\documentclass{csmagclass}														

\begin{document}		
\headings																															 %

\title{{ Magnetism of the diluted Ising antiferromagnet in a magnetic field on the Kagome lattice: single-spin cluster approximation }}
\author[1]{{ A. BOB\'AK \thanks{Corresponding author: andrej.bobak@upjs.sk}}}
\author[1]{{ T. LU\v{C}IVJANSK\'Y}}
\author[1]{{ M. \v{Z}UKOVI\v{C}}}
\affil[1]{{Institute of Physics, Faculty of Science, P. J. \v{S}af\'arik University, Park Angelinum 9, 

041 54 Ko\v{s}ice, Slovak Republic}}

\maketitle

\begin{Abs}
An effective-field theory based on the single-spin cluster has been used to study of a diluted spin-1/2 Ising antiferromagnet on the Kagome lattice with nearest-neighbor interactions. We observe five plateaus in the magnetization curve of the diluted antiferromagnet when a magnetic field is applied which is in agreement with Monte Carlo calculation. The effect of the site-dilution on the magnetic susceptibility is also investigated and discussed. In particular, we have found that the frustrated Kagome lattice inverse susceptibility fall to zero at $0$ K. 
\end{Abs}
\keyword{geometrical frustration, Kagome lattice, effective-field theory }
\section{Introduction}
\hspace*{0.5cm}It has long been known that frustration due to lattice geometry in some Ising systems can result in infinite ground-state degeneracy and no long-range order at any temperature. The simplest model of geometrical frustration is the triangle composed of three spins on the vertexes with the antiferromagnetic (AFM) interaction between them. Examples of the two-dimensional crystal lattices composed of these triangles are the triangular and Kagome lattices. Namely, in the latter kind of lattice, the frustrated triangles are arranged to share sites (corner-sharing triangles, Fig.~1) instead of bonds (edge-sharing triangles) as is the case in the former lattice. As has pointed out in \cite{ref1}, the AFM system in Kagome lattice shows huge degeneracy of ground states and small magnetic field $h$ can lift this degeneracy partially. For this reason in an effective-field theory two magnetization plateaus are formed for $h/|J|<4.0$ \cite{ref2}. This finite-field macroscopic degeneracy is not present in the frustrated triangular lattice which shows, by using the same approximate theory, only one plateau with magnetization $m =1/3$ \cite{ref3}. Therefore, it is especially important in approximate theories to incorporate possible perturbations in this Kagome AFM system to lift this degeneracy and try to accurately describe this system, when compared with both Monte Carlo and experimental data \cite{ref4}. \\
\begin{figure}[h!]
\centering
\includegraphics[width=6cm]{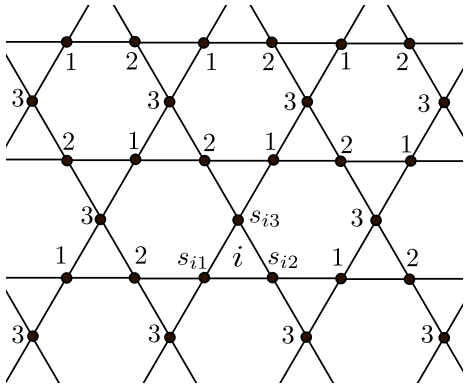}
\caption{Illustration of the Kagome lattice divided into three sublattices 1, 2, and 3 (see text).}
\label{fig1}
\end{figure}
\hspace*{0.5cm}The purpose of the present work is to study effects of site dilution by nonmagnetic impurities of the otherwise perfect Kagome lattice on the magnetic properties with paying attention to the magnetic susceptibility. This problem is relevant for establishing comparisons with experimental data as, obviously, any real material contains a certain amount of dilution. These systems are the most simple disordered materials because the perfect crystallographic lattice exists in it, with the lattice sites occupied at random by magnetic or nonmagnetic atoms. To our knowledge no analytical expression, exact or approximate, which is valid at all temperatures, has so far been proposed for such disordered AFM Kagome lattice.         
%
\section{Model and formalism}
\hspace*{0.5cm}Let us introduce the Hamiltonian for a site-dilute Ising AFM, treated in this work, in the following form  
\begin{eqnarray}
\label{Hamil}
H= -J\sum_{(i,j)}s_is_j\xi_i\xi_j - h\sum_{i}s_i\xi_i,
\end{eqnarray}
where $J<0$ represents the AFM interaction, $h$ is the magnetic field, $s_i$ are the Ising spins $(s_i\pm 1)$, and $(i,j)$ denotes the summation over all the nearest-neighboring spin pairs. $\{\xi_j\}$ is a set of independent, uniformly distributed random variables which take values of unity or zero, depending on whether the site $j$ is occupied by a magnetic atom or not.\\
\hspace*{0.5cm}However, like in the triangular lattice \cite{ref3}, to include the effect of the geometrical frustration, we chose to partition the Kagome lattice into three interpenetrating sublattices $1, 2$ and $3$ in such a way that spins on one sublattice only interact with spins from the other two sublattices (see Fig.~1). Now, by following the same procedure as that in Ref. \cite{ref3}, the expressions for the averaged sublattice magnetizations $m_\nu = \langle \langle s_{i\nu} \rangle_0 \xi_i \rangle_c, (\nu =1, 2, 3)$, where $\langle ... \rangle _0$ ($\langle ...\rangle_c$) indicates the thermal (configurational) average, are given by
\begin{equation}
\label{mag1}
m_1 = p (a + m_2 b)^2(a + m_3 b)^2 \tanh[\beta(x+h)]|_{x = 0},
\end{equation}
\begin{equation}
\label{mag2}
m_2 = p (a + m_1 b)^2(a + m_3 b)^2 \tanh[\beta(x+h)]|_{x = 0},
\end{equation}
\begin{equation}
\label{mag3}
m_3 = p (a + m_1 b)^2(a + m_2 b)^2 \tanh[\beta(x+h)]|_{x = 0},
\end{equation}
with $a = 1-p + p\cosh(JD)$, $b = \sinh(JD)$, $\beta =1/k_BT$, where $p$ is the concentration of magnetic atoms defined by $p = \langle\xi_j\rangle_c$. The explicit form of Eqs. (2)-(4) can
be calculated by using the mathematical relation $\exp(\alpha D) f(x) = f(x+\alpha)$, where $D = \partial/\partial x$ is the differential operator. Then we can define the total magnetization per site $m = (m_1 + m_2 + m_3)/3$ from which the total initial susceptibility per site is given by
\begin{equation}
\label{susc}
\chi = \lim_{h \rightarrow 0} \frac{\partial m}{\partial h} = \frac{1}{3}(\chi_1 + \chi_2 + \chi_3),
\end{equation}
where $\chi_\nu$ is the sublattice initial susceptibility defined by $\chi_\nu =\lim_{h \rightarrow 0} (\partial m_\nu /\partial h)$, $(\nu = 1, 2, 3)$. 
\section{Results and discussion}
\hspace*{0.5cm}In zero magnetic field, Eqs. (\ref{mag1})-(\ref{mag3}) have only the solution  with $m = 0$ at all temperatures, which means that our effective-field approach reproduces the exact result of no long-range order down to $T=~0$~K~\cite{ref5}. \\
\begin{figure}[h!]
\centering
\includegraphics[width=8cm]{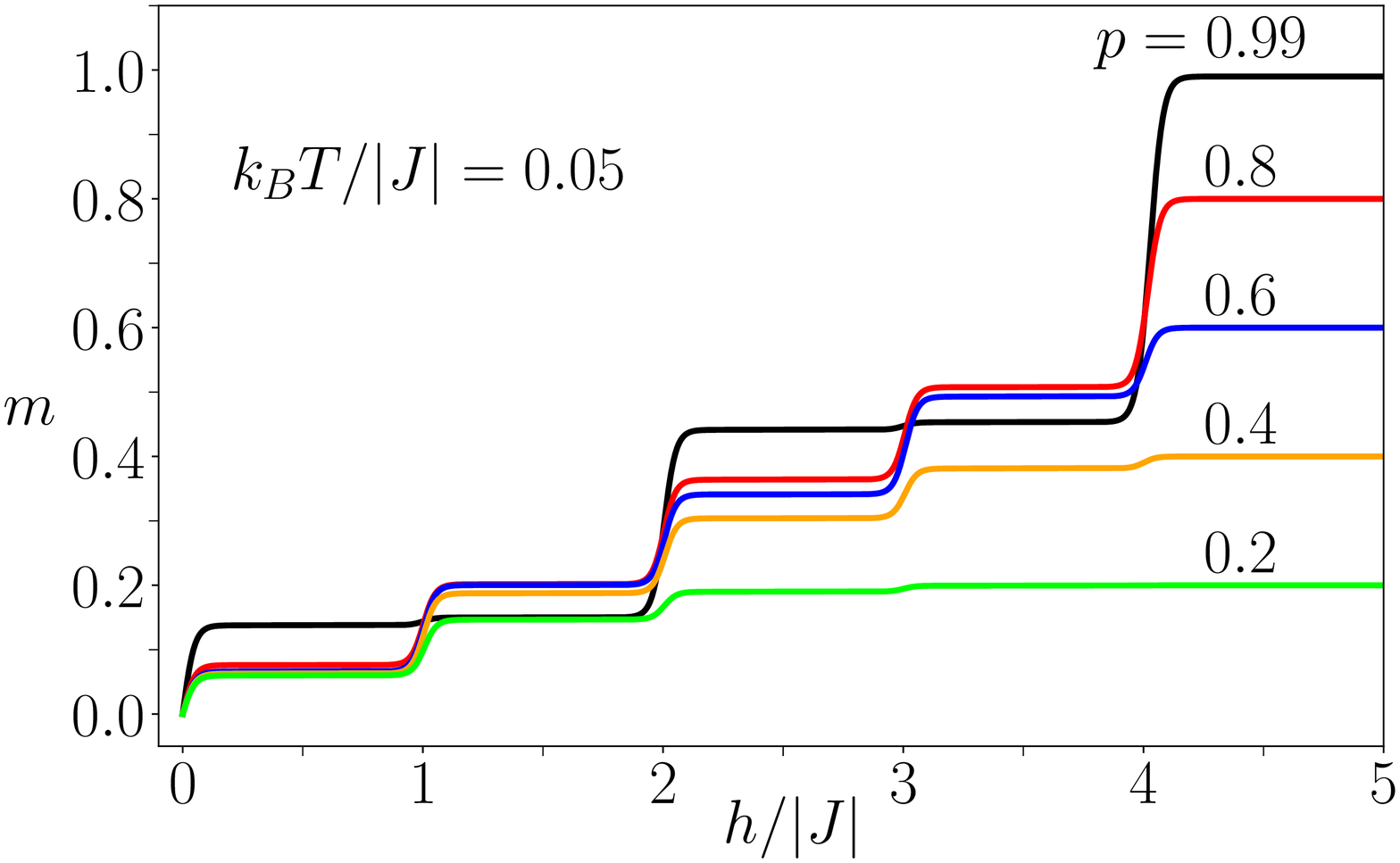}
\caption{Magnetization as function of $h/|J|$ for the diluted AFM Ising model on the Kagome lattice with $k_BT/|J| = 0.05$.}
\label{fig1}
\end{figure}
\hspace*{0.5cm}On the other hand, in Fig.~2 we plot the field dependence of the magnetization at low temperature, namely $k_BT/|J| = 0.05$, for diluted cases ($p < 1$) with different degrees of dilution. In this case, we observe five plateaus in the magnetization curves between integer values of the reduced magnetic field $h/|J|\leq 4.0$. This behavior is in excellent agreement with the Monte Carlo method \cite{ref6}, with the saturated value of the magnetization equal to $p$. This situation contrasts with pure case (or $p=1.0$), where two unphysical plateaus are observed for $h/|J|<4.0$ \cite{ref2}. We note that in the case of the diluted Kagome lattice, the values of the crossover magnetic fields $h_c/|J|$ can be obtained by considering the energy of two corner-sharing triangles. We will not repeat such calculations here, as they can be found in Ref. \cite{ref6}, where the origin of the five magnetization plateaus in Fig.~2 is also elucidated. \\
\begin{figure}[h!]
\centering
\includegraphics[width=8cm]{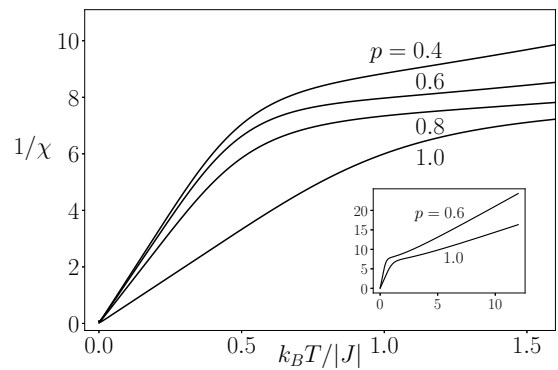}
\caption{The zero-field inverse susceptibility per spin for the diluted AFM Ising model on the Kagome lattice versus temperature for different concentrations of magnetic atoms. The inset shows two curves of the same quantity including very high temperatures.}
\label{fig1}
\end{figure}
\hspace*{0.5cm}As can be seen from Fig.~3, the zero-field inverse susceptibility for different concentrations of magnetic atoms in our AFM Ising model on the Kagome lattice goes to the zero for $T\rightarrow 0$ K. On the other hand, the inverse susceptibilities at high temperatures correspond to the linear Curie-Weiss law. In order to see this linear behavior more clearly, we show in the inset of the figure examples of two curves with the concentrations $p=1.0$ and $p=0.6$ including very high temperatures. Generally, the asymptotes of these inverse susceptibilities pass through the temperature axis at the negative Curie-Weiss temperature $\Theta$. From the detailed numerical calculations we find that for $p =1.0$ (undiluted case) the value of $k_B|\Theta|/|J|= 4.7384$,  which  decreases almost linearly to the zero when $p\rightarrow 0$. This fitted curve of the Curie-Weiss temperature $\Theta$ is shown in Fig.~4. We note that for $p=0.0$ the Curie-Weiss temperature vanishes (empty circle), since in the absence of magnetic atoms the susceptibility should correspond to that of the nonmagnetic Kagome lattice.   
%
%

%
%
%
\begin{figure}[h!]
\centering
\includegraphics[width=7cm]{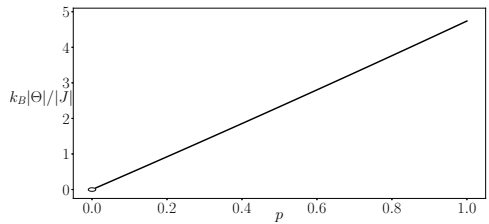}
\caption{The fitted curve of the concentration dependence of the Curie-Weiss temperature $k_B|\Theta|/|J|$ for the diluted AFM Ising model on the Kagome lattice.}
\label{fig1}
\end{figure}
\section{Conclusions}
\hspace*{0.5cm}We have used the effective-field in order to study magnetic ordering on the highly frustrated diluted Kagome lattice. The calculation demonstrates that the AFM Ising model on the Kagome lattice due to the frustration does not show long-range order even at $T= 0$~K, which is exact. However, a small magnetic field applied together with the site dilution can lift this degeneracy and the resulting magnetization exhibits five plateaus in contrast with the case of the pure model, which displays, for $h/|J|<4.0$, two unphysical plateaus \cite{ref2}. Therefore, we have shown that in the analytical but approximate theory it is important to incorporate another perturbation to accurate describe such highly frustrated system. In our case the correct theoretical results come from the interplay of dilution and magnetic field.  \\
\hspace*{0.5cm}Another effect of dilution is related to the inverse susceptibility in the whole temperature range. We have found that both the nondiluted and diluted frustrated Kagome lattice inverse susceptibilities fall to zero at $0$ K. Thus, there is a divergence of the susceptibility as we approach $T = 0$ K for all the cases considered in this work. Even though we will not present a comparison with experimental measurements here, it is interesting to note that some of the features exhibited by the calculated susceptibility are qualitatively similar to those experimentally found in the iron jarosite \cite{ref7} which is considered as an experimental realization of the Kagome lattice.   \\ 

%

\section{Acknowledgement}
This work was supported by the Scientific Grant Agency of Ministry of Education of Slovak Republic (VEGA No. 1/0531/19) and the scientific grant of Slovak Research and Development Agency (Grant No. APVV-16-0186).
%


\end{document}